\documentclass
[preprintnumbers,preprint,tightenlines,prd,amsmath,amssymb,nofootinbib]
{revtex4-1}

\usepackage{xspace}
\usepackage[utf8]{inputenc}
\usepackage{graphicx}
\usepackage[tight]{subfigure}

\newcommand{\order}{\mathop{\mathcal{O}}\nolimits}
\newcommand{\lagrangian}{\mathcal{L}}
\newcommand{\mh}{m_h}
\newcommand{\mhcentval}{125.09\,\GeV}
\newcommand{\mt}{m_t}
\newcommand{\tanb}{\tan\!\beta}
\newcommand{\cotb}{\cot\!\beta}

\newcommand{\cosb}{\cos\!\beta}
\newcommand{\mA}{m_A}
\newcommand{\MSUSY}{M_{\mathrm{SUSY}}}
\newcommand{\MS}{\widetilde{m}}
\newcommand{\Monehalf}{m_{1/2}}
\newcommand{\MGUT}{M_{\mathrm{GUT}}}

\newcommand{\GeV}{\mathrm{GeV}}
\newcommand{\TeV}{\mathrm{TeV}}
\newcommand{\paper}{article\xspace}

\newcommand{\MSbar}{\ensuremath{\overline{\text{MS}}}\xspace}
\newcommand{\DRbar}{\ensuremath{\overline{\text{DR}}}\xspace}
\newcommand{\scale}{Q}
\newcommand{\gYMSSMDR}{g_Y}

\newcommand{\gtwoMSSMDR}{g_2}

\newcommand{\ybSMMS}{y_b}
\newcommand{\ybMSSMDR}{h_b}
\newcommand{\ytauSMMS}{y_\tau}
\newcommand{\ytauMSSMDR}{h_\tau}
\newcommand{\ytMSSMDR}{h_t}
\newcommand{\asSMMS}{{\alpha}_s}

\newcommand{\mgluino}{M_3}

\newcommand{\Tt}{T_t}
\newcommand{\At}{A_t}
\newcommand{\Xt}{X_t}
\newcommand{\Tb}{T_b}
\newcommand{\Ab}{A_b}
\newcommand{\Xb}{X_b}

\newcommand{\Atau}{A_\tau}
\newcommand{\Xtau}{X_\tau}
\newcommand{\Gammavac}{\Gamma_\mathrm{vac}}

\newcommand{\modelname}[1]{\texttt{#1}\@\xspace}
\newcommand{\code}[1]{\texttt{#1}\@\xspace}
\newcommand{\FS}{\texttt{FlexibleSUSY}\@\xspace}
\newcommand{\feft}{\texttt{Flex\-ib\-le\-EFT\-Higgs}\@\xspace}
\newcommand{\susyhd}{\texttt{SusyHD}\@\xspace}
\newcommand{\softsusy}{\texttt{SOFTSUSY}\@\xspace}
\newcommand{\sarah}{\texttt{SARAH}\@\xspace}
\newcommand{\HSSUSY}{\modelname{HSSUSY}}
\newcommand{\SplitMSSM}{\modelname{SplitMSSM}}
\newcommand{\CT}{\texttt{CosmoTransitions}\@\xspace}

\begin{document}

\preprint{KIAS--Q18019}
\title{Higgs mass and vacuum stability with high-scale supersymmetry}
\author{Jae-hyeon~Park}
\affiliation{Quantum Universe Center,
  Korea Institute for Advanced Study,
  85 Hoegiro Dongdaemungu,
  Seoul 02455, Republic of Korea}

\begin{abstract}
In the high-scale (split) MSSM,
the measured Higgs mass sets an
upper bound on the supersymmetric scalar mass scale $\MSUSY$
around $10^{11}$ ($10^{8}$) GeV,
for $\tanb$ in the standard range and the central value of
the top quark mass $\mt$.
This \paper discusses how maximal $\MSUSY$ is affected by
negative threshold corrections to the quartic Higgs coupling
arising from the sbottom and stop trilinear couplings.
In the high-scale MSSM with very high $\tanb$,
the electroweak vacuum decay due to the large bottom Yukawa coupling
rules out the possibility of raising $\MSUSY$ beyond the above limit.
In cases with large $\Ab$ or $\At$,
$\MSUSY$ as a common mass of the extra fermions and scalars
can be as high as $10^{17}\,\GeV$
remaining consistent with $\mh$ and the vacuum longevity
if $\mt$ is smaller than the central value by $2\sigma$.
For the central value of $\mt$,
the upper limit on $\MSUSY$ does not change very much
owing to the metastability,
which is the case also in the split MSSM
even with $\pm 2\sigma$ variations in $\mt$.
\end{abstract}
\maketitle







The Large Hadron Collider is giving continuous blows to
the idea of natural supersymmetry,
its discovery as a major objective of the machine notwithstanding.
The pressures are both direct and indirect.
Direct searches for the supersymmetric particles are
pushing the mass bounds up
\cite{SUSY searches}.
The measured Higgs mass is hinting indirectly at
the stop mass scale
(possibly orders of magnitude) higher than $\order(10)\,\TeV$
\cite{mh mstop}.
Nevertheless, supersymmetry remains one of the most elegant
frameworks in which to structure a fundamental theory of the nature.
The existence of supersymmetry, if not at the TeV scale,
would be more plausible
if the superstring theory is assumed to be the quantum mechanical
description of gravity.
Supersymmetry at high scales might also be motivated
from model-building perspectives,
for instance as a setting for
EeV-scale gravitino dark matter \cite{EeV Gravitino DM},
or as a selection out of the string landscape \cite{supersplit}.
This thought prompts the question of
how high the supersymmetry scale might be.

Shortly after the Higgs mass $\mh$ was measured,
it was pointed out within the Standard Model (SM)
that renormalization group running drives
the quartic Higgs coupling $\lambda$ negative
above a scale around $10^{10}\,\GeV$,
if the top quark mass $\mt$,
the strong coupling $\asSMMS$, and
$\mh$ are taken to be their central values
\cite{Degrassi:2012ry}.
This implies that high-scale supersymmetry is disfavoured
if $\MSUSY$, the mass scale of the supersymmetric particles
including the extra Higgses,
is significantly higher than the zero of $\lambda(\scale)$
\cite{Bernal:2007uv,Giudice:2011cg,Bagnaschi:2014rsa,Allanach:2018fif}.
In the effective field theory (EFT) formalism
with the minimal supersymmetric standard model (MSSM)
as the full theory,
this is due to the matching condition,
\begin{equation}
  \label{eq:lambda matching}
  \lambda = \frac{1}{4} (\gtwoMSSMDR^2 + \gYMSSMDR^2) \cos^2 2\beta +
            \Delta\lambda,
\end{equation}
where the leading term, given by
squares of gauge couplings and cosine of the Higgs mixing angle $\beta$,
is non-negative,
and $\Delta\lambda$ is the threshold correction.
As the matching scale, chosen to be $\MSUSY$, grows so high that
$\lambda$ becomes negative,
the above condition becomes difficult to fulfil
unless $\Delta\lambda$ is large negative enough.

Proposed ways to circumvent the upper limit on $\MSUSY$ can be
classified into the following three categories:
(a) allowing for large enough uncertainties in
low energy parameters such as $\mt$ and $\mh$
for $\lambda$ to stay non-negative up to the Planck scale
\cite{Ibe:2013rpa},
(b) extending the low energy EFT to make the matching condition
easier to satisfy
\cite{Benakli:2013msa,Bagnaschi:2015pwa},
or
(c) realizing a large negative threshold correction
\cite{Vega:2015fna,Ellis:2017erg}.
For the last option,
Ref.~\cite{Ellis:2017erg} considered significantly non-degenerate
spectra of the supersymmetric particles.
In this way, even a unification-scale $\MSUSY$ has been shown to be viable
provided that the Higgsino mass $\mu$ is much lower than
the common gaugino mass $\Monehalf$ to the extent that
$\mu/\Monehalf \sim 10^{-4}$.
This mass hierarchy would imply that $|\mu| \lesssim 10^{14}\,\GeV$
as long as the heaviest particle has a sub-Planckian mass.
In \cite{Vega:2015fna} on the other hand,
$\tanb$ was pushed up to very high values such that
a bottom Yukawa coupling larger than unity
can be a source of $\Delta\lambda$ that can overcome
the tree-level part of $\lambda$.
The sbottom threshold correction is enhanced
due to the large coupling that multiplies
the scalar trilinear interaction term,
\begin{equation}
  \label{eq:yb mu trilinear}
  \Delta\lagrangian_F = \ybMSSMDR\mu H_u \widetilde{Q}^* \tilde{b}_R .
\end{equation}
A Landau pole around $10 \MSUSY$ stemming from
a large but still perturbative value of
the bottom Yukawa coupling $\ybMSSMDR$ \cite{Vega:2015fna},
may not be regarded as a critical flaw of the scenario.
As mentioned in that reference, however,
what is not clear is whether the electroweak vacuum would be
stable enough or not.

In this \paper,
the possibility of fitting $\mh$ shall be contemplated
in the context of high-scale (split) supersymmetry
using the above supersymmetric as well as
the following soft supersymmetry breaking trilinear terms,
\begin{equation}
  \Delta\lagrangian_\mathrm{soft} =
  - \Tb H_d \widetilde{Q} \tilde{b}_R^*
  + \Tt H_u \widetilde{Q} \tilde{t}_R^* ,
\end{equation}
wherein
it is common to factor the MSSM Yukawa couplings
$\ybMSSMDR$ and $\ytMSSMDR$ out of the $T$-parameters,
yielding the familiar definitions of the $A$-parameters:
\begin{equation}
  \Tb \equiv \ybMSSMDR \Ab,
  \quad
  \Tt \equiv \ytMSSMDR \At.
\end{equation}
To express contributions to $\Delta\lambda$,
it is common to define the left-right squark mixing parameters,
\begin{equation}
  \Xb \equiv \Ab - \mu\tanb,
  \quad
  \Xt \equiv \At - \mu\cotb.
\end{equation}
One could also use the trilinear couplings of staus
instead of sbottoms, as the origins of negative threshold corrections.
The results would then be similar to those using sbottoms.
Throughout this \paper, it shall be assumed that
$\Xtau \equiv A_\tau - \mu\tanb = 0$.

As large trilinear couplings can cause charge/color breaking (CCB)
global minima \cite{ccb},
the vacuum metastability shall be required as follows:
\begin{equation}
  \label{eq:metastability condition}
  (\Gammavac/V)\,T^4 < 1 ,
\end{equation}
where $\Gammavac/V$ is
the decay rate of the electroweak vacuum per unit volume
and $T$ is the age of the Universe.
In the semiclassical formulation \cite{tunnelling},
the vacuum decay rate reads
\begin{equation}
  \label{eq:Gamma over V}
  \Gammavac/V = A\,\exp( -S [\overline{\phi}] ) ,
\end{equation}
where $A$ is a prefactor of mass dimension 4,
$S$ is the Euclidean action, and
the ``bounce'' $\overline{\phi}$
is an O(4)-symmetric stationary point of $S$
\cite{Coleman:1977th}.
Thanks to this O(4)-symmetry, $S$ can be put into the form,
\begin{equation}
  \label{eq:action}
 S[\phi(\rho)] = 2 \pi^2 \int_0^\infty d\rho \rho^3
 \left[ \frac{1}{2}\!\left(\frac{d\phi}{d\rho}\right)^2 + V(\phi) \right] .
\end{equation}
The boundary conditions on $\overline{\phi}(\rho)$ then become
\begin{equation}
 \overline{\phi}(\rho \rightarrow \infty) = \phi_+, \quad
 \frac{d\overline{\phi}}{d\rho} (\rho=0) = 0 ,
\end{equation}
where $\phi_+$ denotes the false vacuum.
The bounce $\overline{\phi}(\rho)$ is found numerically
using the \CT package
\cite{Wainwright:2011kj}.
The Euclidean action $S [\overline{\phi}]$ thus obtained
has been compared with that from another numerical method
described in \cite{Park:2010rh}
at selected points in the parameter space,
resulting in good agreement.

The tree-level MSSM scalar potential is substituted for
$V(\phi)$ in \eqref{eq:action} with the restricted set of fields,
\begin{equation}
  \phi =
  \{ h, H, \tilde{b}_L, \tilde{b}_R \}
  \quad\text{or}\quad
  \{ h, H, \tilde{t}_L, \tilde{t}_R \},
\end{equation}
where the enumerated elements are
the SM-like lighter and the heavier $CP$-even Higgses,
as well as the real parts of
the left- and the right-handed sbottoms or stops,
respectively.
Either set is chosen depending on whether
a sbottom or stop trilinear coupling is responsible for
the enhancement of $\Delta\lambda$.
The real parts of the squarks in the above sets are normalized
as real scalar fields.
%

The prefactor $A$ in \eqref{eq:Gamma over V}
shall be estimated to be $(\MSUSY)^4$
on dimensional grounds.
Methods have been developed to calculate $A$
at one-loop level which reduce
the renormalization scale dependence of $\Gammavac/V$
\cite{Endo:2015ixx}.
As the running parameters determine $S$ which
in turn is exponentiated in \eqref{eq:Gamma over V},
the uncertainty in $\Gammavac/V$ from the scale dependence
is indeed exponentially amplified.
Conversely, this implies that
the limits on the trilinear couplings
from \eqref{eq:metastability condition} depend
on the scale only logarithmically.
Therefore, the above simple-minded estimate of $A$
should be enough at least to understand qualitatively
the impact of metastability on $\mh$ from high-scale supersymmetry.
Another issue with the calculation of $\Gammavac/V$
is its gauge dependence which has also been addressed
\cite{Endo:2017gal}.
It should be a meaningful future project to improve the present analysis
resolving the scale and gauge dependence.

The bounce action $S [\overline{\phi}]$ is classically invariant
under scaling of the parameters in the scalar potential $V(\phi)$
\cite{Claudson:1983et}.
Therefore, a large dimensionless ratio $X_{b,t}/\MSUSY$ is well capable of
disturbing the vacuum stability no matter how high $\MSUSY$ is,
which happens to be the same ratio that
controls dominant negative contributions to $\Delta\lambda$.
This ``non-decoupling'' property of metastability
has been demonstrated in the context of
flavour physics as a probe of
flavour-violating trilinear couplings
\cite{Park:2010wf}.
An EFT formulation has also been employed to argue that
disturbance to the vacuum lifetime is not simply suppressed
by pushing up the new physics scale \cite{Branchina:2013jra}.


The SM-like Higgs mass $\mh$ is computed in the EFT approach using
\FS \cite{Athron:2014yba} version 2.2.0 \cite{Athron:2017fvs}
grown out of \softsusy \cite{softsusy},
in combination with \sarah \cite{sarah}.
The bundled model definitions,
\HSSUSY
\cite{HSSUSY,Degrassi:2012ry,Vega:2015fna,Martin:2014cxa}
and
\SplitMSSM
\cite{Degrassi:2012ry,Martin:2014cxa,Benakli:2013msa},
are used for the high-scale and the split MSSM, respectively,
with the following modifications:
(a)
\SplitMSSM is modified to compute $\mh$
using the \feft method
\cite{SplitMSSMEFTHiggs},
(b)
\SplitMSSM is modified to
include the one-loop sbottom and stau threshold corrections to
$\lambda$, taken from \HSSUSY,
(c) the $\tanb$-enhanced corrections to $\ybMSSMDR$ and $\ytauMSSMDR$
proportional to $g_Y^2$ or $g_2^2$
are added.
These latter corrections are not included in the above model files
as some of the implemented results are in
the ``gaugeless'' limit where
loops controlled by $g_{Y,2}$
are neglected \cite{Bagnaschi:2017xid}.
These gauge couplings become comparable to $g_3$
and may thus be non-negligible
as the renormalization scale approaches the unification scale $\MGUT$.
Details of the modifications are documented in the appendix.

The numerical analysis involves the following MSSM parameters as input:
$\tanb$,
the scalar trilinears $A_{b,t}$,
the Higgsino mass $\mu$,
the $CP$-odd Higgs mass $\mA = \MS$,
the soft sfermion masses
$m^2_{\widetilde{Q}} = m^2_{\tilde{b}_R} = m^2_{\tilde{t}_R} =
m^2_{\widetilde{L}} = m^2_{\tilde{\tau}_R} = \MS^2$,
the gaugino masses $M_{1,2,3} = \Monehalf$.
For a given pair of $(\tanb, A_q)$ with $q = b\text{ or }t$,
$\MSUSY = \MS$ is found such that $\mh = \mhcentval$
\cite{Aad:2015zhl}
as is calculated by \FS taking $\MSUSY$ as the matching scale.
Further assumptions about the remaining mass parameters are:
$|\mu| = \Monehalf = \MSUSY$
in the high-scale MSSM, and
$\mu = \Monehalf = 1\,\TeV$
in the split MSSM\@.
To isolate the effect of each trilinear coupling,
either of $X_{b,t}$ is fixed at zero when it is not being scanned.
At each such $\MSUSY$, $\tanb$ or $T_q$ is subsequently varied,
with $\MSUSY$ fixed,
until the left-hand side of \eqref{eq:metastability condition}
becomes unity, yielding the metastability bound.
For this, the additional MSSM parameters
$
g_Y, g_2, g_3, h_{b,t} 
$,
at the scale $\MSUSY$ in the \DRbar scheme,
are put into the tree-level scalar potential
$V(\phi)$ in \eqref{eq:action}
together with $\mu, T_{b,t}$ already specified above.
The \DRbar gauge and Yukawa couplings are obtained
from the \MSbar couplings output by \FS
using the conversion formulae available from the \susyhd package
\cite{Vega:2015fna}.
The soft mass parameters $B_\mu, m^2_{H_d}, m^2_{H_u}$
are determined at tree level by
$\mA$ and the electroweak symmetry breaking conditions.

First, the high-$\tanb$ scenario from \cite{Vega:2015fna} is revisited.
The one-loop sbottom threshold correction to $\lambda$ looks like
\begin{equation}
  \label{eq:Delta lambda 1 sbottom}
  \Delta\lambda^{(1)}_{\tilde{b}} =
  -\frac{(\ybMSSMDR\mu)^4}{32\pi^2\,\MS^4} .
\end{equation}
Making use of this contribution,
one can increase $\tanb$ to the extent that
a large enough MSSM Yukawa coupling $\ybMSSMDR$ allows
\eqref{eq:lambda matching} to hold
for an arbitrarily high $|\mu| = \MS = \MSUSY$.
Note however that the same product $\ybMSSMDR\mu$
affects not only the above threshold correction
but also the existence of CCB global minima
as suggested by \eqref{eq:yb mu trilinear}.
One should therefore pay attention to the stability
of the electroweak vacuum at the same time.

\begin{figure}
  \centering
  \includegraphics{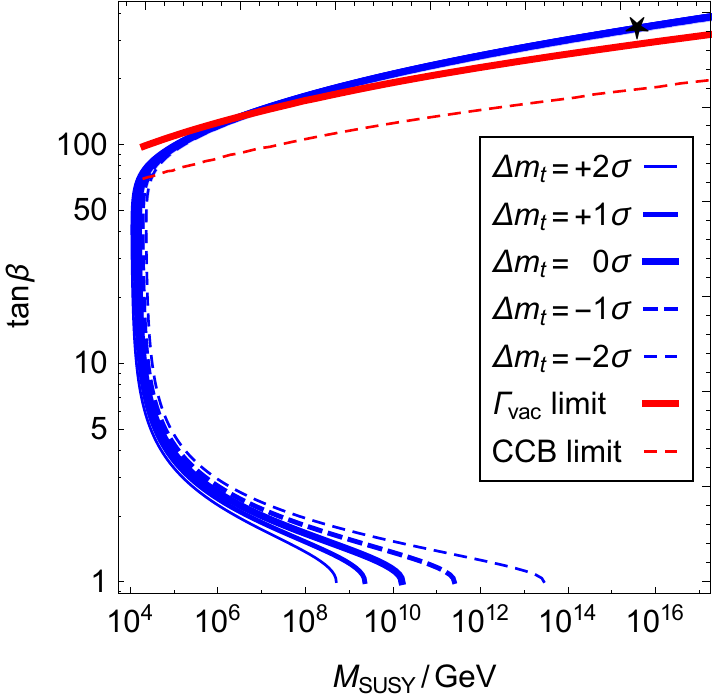}\qquad
  \includegraphics{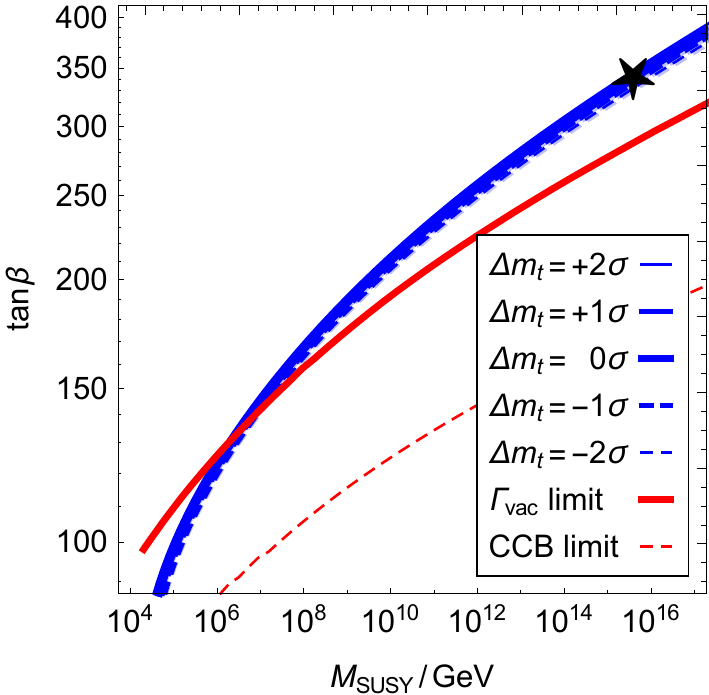}
  \caption{Higgs mass curves (blue) and (meta)stability limit (red) on
    the $(\MSUSY, \tanb)$ plane.  Both plots are the same except that
    the right panel is restricted to the high-$\tanb$ range.
    The thickness and pattern of each blue curve reproducing
    $\mh = \mhcentval$, indicate the size and sign of the deviation
    of used $\mt$ from the central value, respectively.
    The starred point, ruled out by metastability,
    results in the specimen bounce shown in Fig.~\ref{fig:bounce star}.}
  \label{fig:tanb MSUSY}
\end{figure}

The $\mh$ constraint on $\MSUSY$ and $\tanb$ is
reproduced in Fig.~\ref{fig:tanb MSUSY} using \FS.
Both panels are Fig.~6 of \cite{Vega:2015fna} with
$\Xt = \Xtau = 0$ instead of $\At = \MSUSY/2$ and $\Atau = 0$ as well
as the horizontal and vertical axes interchanged.
The sbottom trilinear $\Ab$ is set to zero
and $\mu$ is negative, as in the original plot.
The right panel magnifies the high-$\tanb$ region.
For an estimation of the theory uncertainty due to
missing higher order threshold corrections,
$\lambda$ is matched
at both one- and two-loop levels
by switching the \code{LambdaLoopOrder} parameter \cite{Athron:2017fvs},
leading to the light blue and the blue curves, respectively.
The close proximity of the one-loop matched curves to
the corresponding two-loop matched curves
renders the former hard to see thereby
indicating that
the truncation error is reasonably small.
%
In the region above the red dashed curve,
the scalar potential develops a global CCB minimum
with non-vanishing sbottom vacuum expectation values
due to large $\ybMSSMDR\mu$.
Within that region,
the vacuum longevity condition \eqref{eq:metastability condition}
gives rise to the upper limit on $\tanb$ delineated by the
red solid curve.
It is looser than the CCB limit,
but still excludes the upper part of the blue curves
with $\MSUSY \gtrsim 10^8\,\GeV$.

\begin{figure}
  \centering
  \includegraphics{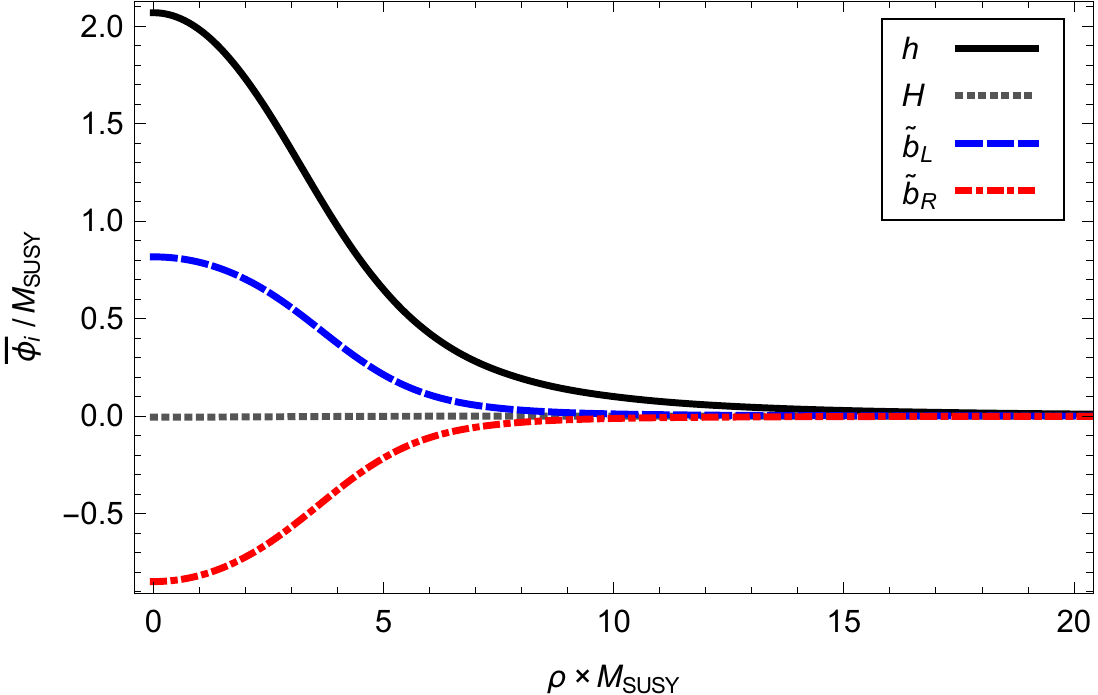}
  \caption{Bounce profile resulting from the starred point in
    Fig.~\ref{fig:tanb MSUSY}, at which
    $\MS = -\mu = \Monehalf = \MSUSY = 3.88 \times 10^{15} \,\GeV$
    and $\tanb = 340$.
    The scalar fields $h, H, \tilde{b}_L, \tilde{b}_R$ are the SM-like
    lighter and the heavier $CP$-even Higgses, as well as the left-
    and the right-handed sbottoms, respectively.
    The real-valued sbottoms are normalized as real scalar fields.}
  \label{fig:bounce star}
\end{figure}%
An instance of the bounce is shown in Fig.~\ref{fig:bounce star} which
corresponds to the point marked with the star in
Fig.~\ref{fig:tanb MSUSY}.
With this bounce as its argument,
the Euclidean action $S [\overline{\phi}]$
evaluates to $275$, much smaller than its lower bound $527$
for the indicated $\MSUSY$.

\begin{figure}
  \centering
  \subfigure[\ High-scale MSSM]{\includegraphics{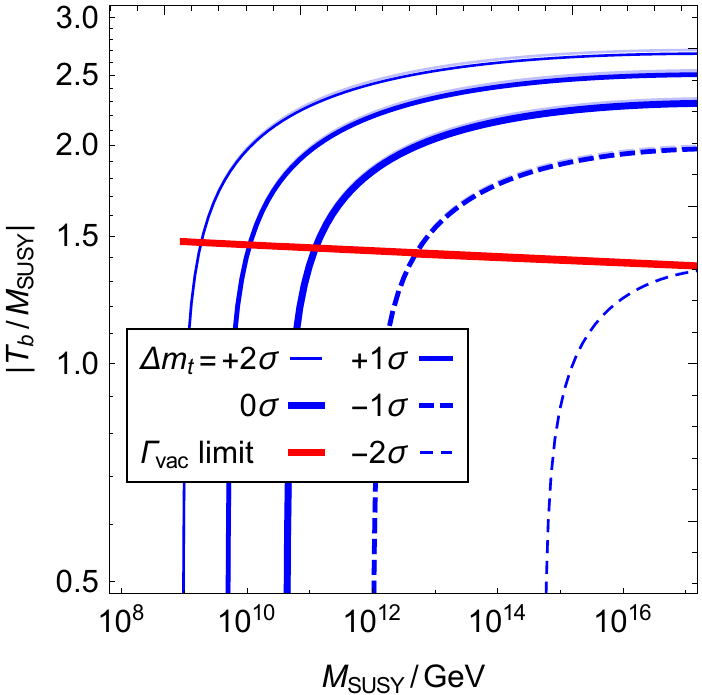}}\qquad
  \subfigure[\ Split MSSM]{\includegraphics{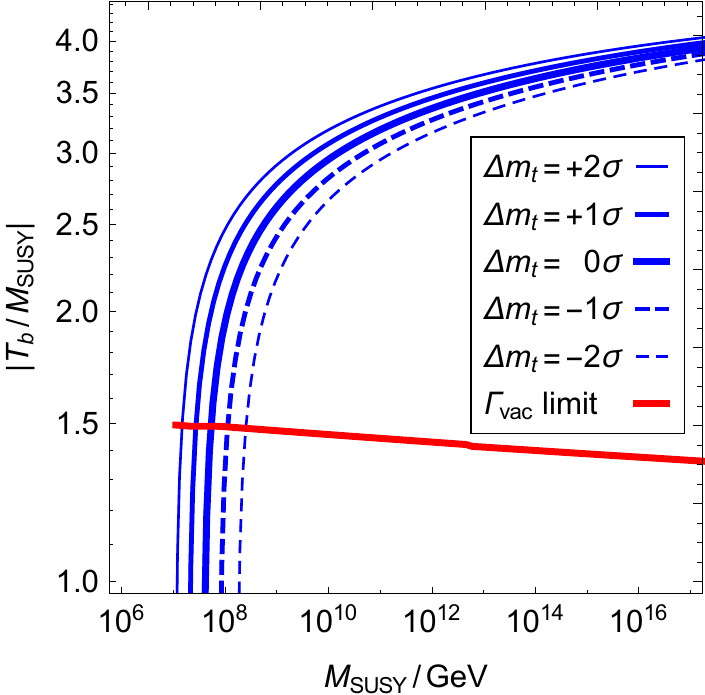}}
  \caption{Sbottom trilinear coupling fitting the Higgs mass (blue) and
    metastability limit (red)
    for $\tanb = 1$,
    as a function of $\MSUSY$ in
    (a) the high-scale MSSM and
    (b) the split MSSM with $\mu = \Monehalf = 1\,\TeV$.
    The thickness and pattern of each blue curve reproducing
    $\mh = \mhcentval$, indicate the size and sign of the deviation
    of used $\mt$ from the central value, respectively.}
  \label{fig:Tb MSUSY}
\end{figure}

An alternative way to enhance negative $\Delta\lambda$ is to
increase $|\Xb|$ far beyond $\sqrt{12}\MSUSY$ via $\Ab$ \cite{Haber:1996fp}.
In this case, one can choose $\tanb$ to be $1$ to minimize the tree-level
contribution to $\lambda$ in \eqref{eq:lambda matching} and
then fit $\mh$ by varying $\Ab$
in the high-scale and the split MSSM\@.
The results are shown in Fig.~\ref{fig:Tb MSUSY},
for $\mu > 0$ and $\Ab < 0$.
The other sign combinations lead to similar outcomes.
The vertical axis is chosen to involve $\Tb$
instead of $\Xb$ or $\Ab$ to avoid displaying huge numbers.
There exists indeed a value of $\Tb$ yielding correct $\mh$
at any $\MSUSY$ within the selected range.
However, the metastability puts a stringent constraint on $\Tb$,
thereby resulting in the limits
(a) $\MSUSY \lesssim 10^{11}\,\GeV$ in the high-scale MSSM and
(b) $\MSUSY \lesssim 10^8\,\GeV$ in the split MSSM,
for the central value of $\mt = 173.34\,\GeV$
\cite{ATLAS:2014wva}.
%
%
%
%
If $\mt$ is shifted from this by $-2\sigma$
with $\sigma = 0.76 \,\GeV$ \cite{ATLAS:2014wva},
one can find solutions for $\MSUSY \gtrsim \MGUT$
allowed by both $\mh$ and the vacuum lifetime in the high-scale MSSM\@.
In the split MSSM by contrast,
varying $\mt$ by $\pm 2\sigma$
does not change the upper limit on $\MSUSY$ very much.
As in Fig.~\ref{fig:tanb MSUSY},
CCB bounds could also be plotted,
which leave $|\Tb/\MSUSY| \lesssim 0.04$ or narrower ranges.
Their boundaries would therefore lie
much lower than the displayed region.

\begin{figure}
  \centering
  \subfigure[\ High-scale MSSM]{\includegraphics{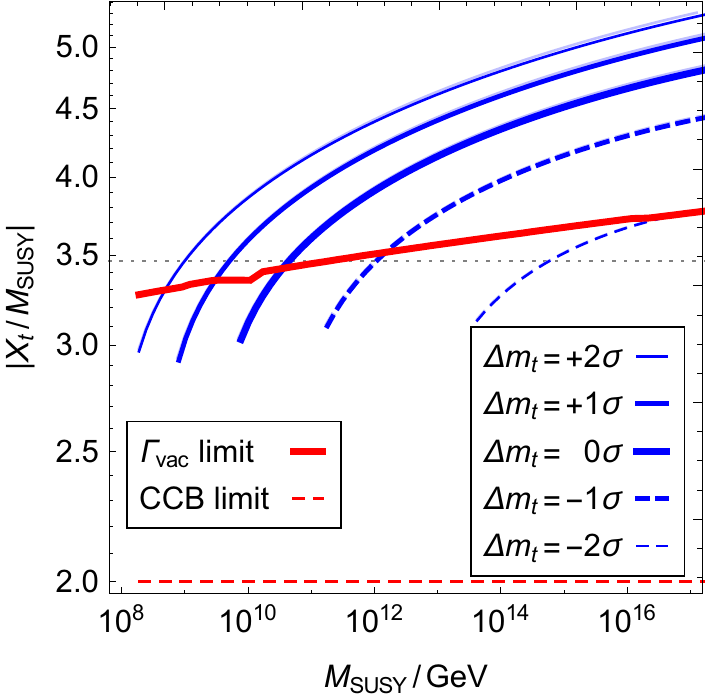}}\qquad
  \subfigure[\ Split MSSM]{\includegraphics{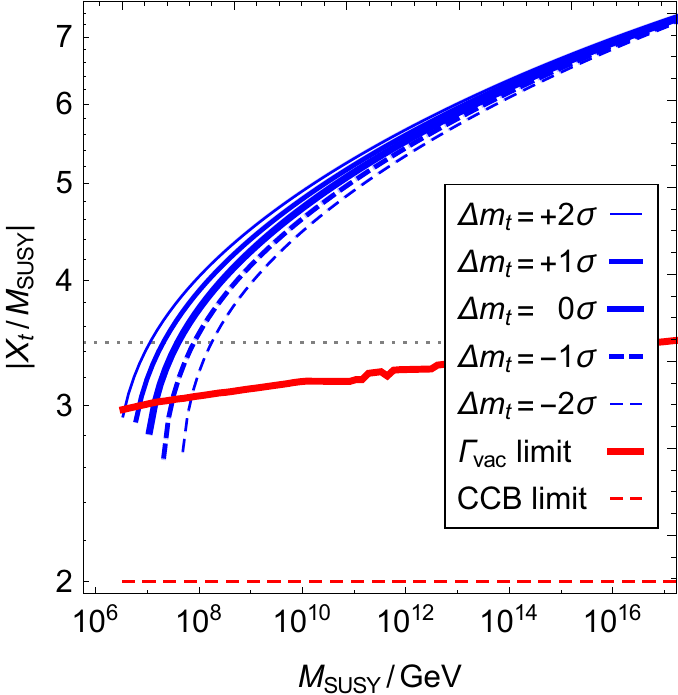}}
  \caption{Stop trilinear coupling fitting the Higgs mass (blue) and
    (meta)stability limit (red)
    for $\tanb = 1$,
    as a function of $\MSUSY$ in
    (a) the high-scale MSSM and
    (b) the split MSSM with $\mu = \Monehalf = 1\,\TeV$.
    The thickness and pattern of each blue curve reproducing
    $\mh = \mhcentval$, indicate the size and sign of the deviation
    of used $\mt$ from the central value, respectively.
    The horizontal dotted line marks the height
    $|\Xt/\MSUSY| = \sqrt{12}$.}
  \label{fig:Xt MSUSY}
\end{figure}

A negative threshold correction can also arise from
$|\Xt| \gtrsim \sqrt{12}\MSUSY$ \cite{Haber:1996fp}.
Values of $\Xt$ leading to correct $\mh$ and the vacuum decay limits thereon
are shown in Fig.~\ref{fig:Xt MSUSY},
for $\mu > 0$ and $\Xt < 0$.
The other sign combinations lead to similar outcomes.
To suppress the tree-level term in \eqref{eq:lambda matching},
$\tanb$ is fixed at $1$.
In the range $0 < |\Xt| < \sqrt{12} \MSUSY$,
the stop threshold correction to $\Delta\lambda$ tends to be positive
with the maximum around $|\Xt| \approx \sqrt{6} \MSUSY$,
thereby pushing $\MSUSY$ down even lower than the value for $\Xt = 0$.
For the purpose of raising $\MSUSY$,
$\Xt$ is therefore better chosen to be zero
rather than a nonzero value in this range.
A large enough $|\Xt|$ on the other hand can yield correct $\mh$
for an arbitrarily high $\MSUSY$.
In the high-scale MSSM,
the metastability bound however happens to range in the vicinity of
$|\Xt / \MSUSY| = \sqrt{12}$ with a crossing point
around $\MSUSY \approx 2 \times 10^{11}\,\GeV$.
This means that the limit $\MSUSY \lesssim 10^{11}\,\GeV$
\cite{Bagnaschi:2014rsa,Allanach:2018fif}
resulting from vanishing $\Xt$ and the central value of $\mt$,
does not change very much
even if $\Xt$ is allowed to be nonzero.
A viable parameter region with $\MSUSY \gtrsim \MGUT$
can still be opened up
if $\mt$ is lowered by $2\sigma$ from the central value.
Note that this ``low-$\mt$ region'' would be
excluded leading to the restriction,
$\MSUSY \lesssim 6 \times 10^{14}\,\GeV$,
even for $\Delta m_t = -2 \sigma$,
if the more stringent CCB bound were adopted instead.
In the split MSSM by contrast,
the vacuum decay excludes all parts of the blue curves with
$|\Xt| > \sqrt{12} \MSUSY$.
Therefore, the resulting maximal $\MSUSY$ for any shown $\mt$
is the same as the corresponding upper limit for $\Xt = 0$.


To sum up,
this \paper has attempted to address the question of
how high the supersymmetry scale can be,
given the measured SM-like Higgs mass
as a constraint on the parameters of the MSSM\@.
Two types have been considered
as to the mass spectrum of the supersymmetric particles
including the extra Higgses:
(a) nearly degenerate fermions and scalars
(high-scale MSSM),
(b) split spectrum where the fermions are at the TeV scale
(split MSSM).
To satisfy
the matching condition on the quartic Higgs coupling,
the sbottom and the stop trilinear couplings
have been employed as sources of
the potentially large negative threshold corrections.
In all cases, it is possible to reproduce
$\mh = \mhcentval$, by choosing appropriate values of
$\tanb$ in combination with $\Ab$ or $\At$,
for $\MSUSY$ up to $10^{17}\,\GeV$.
However, the lifetime of the electroweak vacuum
places severe constraints on the trilinear couplings
and mostly brings back the upper limits on
$\MSUSY$ for $\Xb = \Xt = 0$:
$\MSUSY \lesssim 10^{11}\,\GeV
\text{ and }
10^8\,\GeV$,
in the high-scale and the split MSSM, respectively,
for the central value of $\mt$.
Nevertheless,
a small extension of the viable parameter volume could be
achieved via non-vanishing $\Xb$ and/or $\Xt$ in the high-scale MSSM:
$\MSUSY \gtrsim \MGUT$ becomes viable
if $\mt$ is allowed to be smaller than the central value by $2\sigma$.
Note that this smaller $\mt$ still causes $\lambda$ to turn negative
at a scale around $6 \times 10^{14}\,\GeV$
and therefore that $\MSUSY$ higher than this zero of $\lambda$
requires negative threshold corrections.
On the other hand,
the vacuum decay rate rules out the scenario
where very high $\tanb$ reconciles
arbitrarily high $\MSUSY$ with $\mh$
\cite{Vega:2015fna}.
As already mentioned in the introductory part,
these findings still leave the possibilities of
going beyond the assumptions made in this work
in order to lift the restrictions on the scale of supersymmetry.


\vspace{1ex}
The author thanks
Oscar Vives,
Deog Ki Hong,
Pyungwon Ko,
and
Dominik Stöckinger
for the helpful comments.
He also thanks the KIAS Center for Advanced Computation for
providing computing resources through the Abacus system.

\appendix*

\section{Corrections to Yukawa couplings added to \FS model files}

The SM and the MSSM bottom Yukawa couplings,
$\ybSMMS$ and $\ybMSSMDR$,
are related at one-loop order by
\cite{Delta yb}
\begin{equation}
  \ybSMMS =
  \ybMSSMDR \cosb (
  1
  + \Delta_b^{\tilde{\chi}^0}
  + \Delta_b^{\tilde{\chi}^\pm}
  + \Delta_b^{\tilde{g}}
  )
  ,
\end{equation}
where the $\tanb$-enhanced corrections read
\begin{align}
   \frac{{16\pi^2}}{\tanb} \,\Delta_b^{\tilde{\chi}^0} =&
   +
   \frac{1}{9}
   {g_Y^2}
   \frac{X_b}{\tanb} M_1
   I(m^2_{\tilde{b}_L},m^2_{\tilde{b}_R},M_1^2)
   -
  \frac{1}{6}
  {g_Y^2}
  M_1 \mu
  I(M_1^2,\mu^2,m^2_{\tilde{b}_L})
  \nonumber \\
  &-
  \frac{1}{3}
  g_Y^2
  M_1 \mu
  I(M_1^2,\mu^2,m^2_{\tilde{b}_R})
  -
  \frac{1}{2}
  g_2^2
  M_2 \mu
  I(M_2^2,\mu^2,m^2_{\tilde{b}_L})
  ,
  \label{eq:Delta yb neutralino}
\\
  \frac{{16\pi^2}}{\tanb} \,\Delta_b^{\tilde{\chi}^\pm} =&
  -
    {g_2^2} M_2 \mu
    I(M_2^2, \mu^2, m^2_{\tilde{t}_L})
  +
    {\ytMSSMDR^2}
    \Xt \mu
    I(m^2_{\tilde{t}_L},m^2_{\tilde{t}_R},\mu^2)
  ,
  \label{eq:Delta yb chargino}
\\
  \frac{{16\pi^2}}{\tanb} \,\Delta_b^{\tilde{g}^{\phantom{\pm}}} =&
    -
    \frac{8}{3} {g_3^2}
    \frac{\Xb}{\tanb} \mgluino
    I(m^2_{\tilde{b}_L},m^2_{\tilde{b}_R},\mgluino^2)
  ,
  \label{eq:Delta yb gluino}
\end{align}
in terms of
the loop function
\begin{equation}
  I(a,b,c) \equiv
  \frac%
  {a b \ln(a/b) + b c \ln(b/c) + a c \ln(c/a)}%
  {(a - b) (b - c) (a - c)}
  .
\end{equation}
Likewise, the SM and the MSSM tau Yukawa couplings,
$\ytauSMMS$ and $\ytauMSSMDR$, are related by
\begin{align}
  \ytauSMMS =&\
  \ytauMSSMDR \cosb (
  1
  + \Delta_\tau^{\tilde{\chi}^0}
  + \Delta_\tau^{\tilde{\chi}^\pm}
  )
  ,
   \nonumber \\
   \frac{{16\pi^2}}{\tanb} \,\Delta_\tau^{\tilde{\chi}^0} =&
   -
   {g_Y^2}
   \frac{X_\tau}{\tanb} M_1
   I(m^2_{\tilde{\tau}_L},m^2_{\tilde{\tau}_R},M_1^2)
   +
  \frac{1}{2}
  {g_Y^2}
  M_1 \mu
  I(M_1^2,\mu^2,m^2_{\tilde{\tau}_L})
  \nonumber \\
  &-
  g_Y^2
  M_1 \mu
  I(M_1^2,\mu^2,m^2_{\tilde{\tau}_R})
  -
  \frac{1}{2}
  g_2^2
  M_2 \mu
  I(M_2^2,\mu^2,m^2_{\tilde{\tau}_L})
  ,
  \label{eq:Delta ytau neutralino}
\\
  \frac{{16\pi^2}}{\tanb} \,\Delta_\tau^{\tilde{\chi}^\pm} =&
  -
    {g_2^2} M_2 \mu
    I(M_2^2, \mu^2, m^2_{\tilde{\nu}_\tau})
  .
  \label{eq:Delta ytau chargino}
\end{align}
Among the above terms,
those proportional to $\ytMSSMDR^2$ and $g_3^2$, i.e.\
\eqref{eq:Delta yb gluino} and
the last term of \eqref{eq:Delta yb chargino}
are already implemented in \HSSUSY.
For this work,
\HSSUSY and \SplitMSSM have been modified to include
all the above corrections.

\end{document}